# Anomalous conductance response of DNA wires under stretching

*Bo Song,[1] Marcus Elstner,[2] and Gianaurelio Cuniberti[1,a]*

[1] Institute for Materials Science and Max Bergmann Center of Biomaterials,
Dresden University of Technology, D-01062 Dresden, Germany

[2] Institute for Physical and Theoretical Chemistry,
Technical University at Braunschweig, D-38106 Braunschweig, Germany

**ABSTRACT**: The complex mechanisms governing charge migration in DNA oligomers reflect the rich structural and electronic properties of the molecule of life. Controlling the mechanical stability of DNA nanowires in charge transport experiments is a requisite for identifying intrinsic issues responsible for long range charge transfers. By merging density-functional-theory-based calculations and model-Hamiltonian approaches, we have studied DNA quantum transport during the stretching-twisting process of poly(GC) DNA oligomers. During the stretching process, local maxima in the charge transfer integral $t$ between two nearest-neighbor GC pairs arise from the competition between stretching and twisting. This is reflected in local maxima for the conductance, which depend very sensitively on the coupling to the electrodes. In the case of DNA-electrode couplings smaller than $t$, the conductance versus stretching distance saturates to plateau in agreement with recent experimental observations.

[a]E-mail: g.cuniberti@tu-dresden.de

The great interest which DNA attracted in molecular electronic experiments is related to the vision of a truly bottom up electronics at the molecular scale.[1] The very controversial results concerning the conduction of DNA oligomers are ascribable to their internal complexity and their sensitivity to the surrounding environment. It is currently well accepted[2] that the mechanisms of charge migration in DNA wires cannot be addressed independently of their mechanical properties.

Recently, magnetic AFM experiments in the Bustamante group[3] allowed, while independently stretching and twisting single double stranded DNA oligomers, to access many well controlled structural configurations of DNA wires. When stretching a DNA short molecule by pulling one of its ends with a standard AFM tip and having the other end fixed on a surface, the angle $\phi$ between consecutive base pairs reduces from its equilibrium value of 36 degrees. Smaller angle $\phi$ lead to an increased π-π overlap, resulting in larger charge transfer integrals.[6-8] On the other hand, the combined twist-stretch process leads increased inter-base distances d, effectively decreasing the values of the charge transfer integrals, therefore, antagonizing the effect of twist. This may lead to a complex behavior of electronic properties of DNA during the stretching process. Interestingly, a non monotonous behavior of the transport response of stretched DNA has been in fact reported by Cohen et al.[4] and switching and intermittency in DNA break junctions experiments has been observed by Kang et al.[5] As one possible explanation Kang et al. discuss the change in overlap between neighboring bases as a result of a changed DNA configuration during the stretching process, which has been investigated theoretically before,[9] however, only for few geometry snapshots.

In this letter, we investigate the conductance change of DNA along the conformation transition for the stretch-twist process in more detail. Our calculations demonstrate, that the charges transfer integrals and the tunneling current show such a non monotonic dependence for this process, leading to a switching behavior of the electrical response; the resulting intermitting I-d characteristics may explain the recent experiments findings.[4,5] Our numerical calculations are further supported by a minimal analytical solvable model.

For this study, we parameterize a model Hamiltonian from the density functional theory based tight binding (DFTB) method.[10,11] For every geometry snapshot along the stretch-twist path, the effective HOMO-HOMO charge-transfer-integral, t, between two nearest neighbor GC pairs and the effective HOMO level, $\varepsilon$, within one GC pair are calculated from DFTB. Subsequently, we use model Hamiltonian methods based on these parameters to estimate the non-equilibrium current-voltage characteristics for bias voltages of the order of the HOMO-LUMO gap.

In order to calculate the parameters t and $\varepsilon$, we have implemented in DFTB the molecular fragment orbitals method.[12] The whole molecule is then partitioned into several fragments (shown in figure 2), consisting of hydrogen bonded base pairs, thereby neglecting the electronically inactive sugars and phosphate groups. Every fragment molecular orbital (FMO) $\Psi$, which is localized on the base pair, is calculated within DFTB independently, as a linear combination of atomic orbitals, $\Psi_i = \sum_\mu c_{\mu,i} \eta_\mu$. Here $\mu$ is the index of atomic orbitals (AOs) and i is used for FMOs. Furthermore, we can transform the Kohn-Sham Hamiltonian from the AO picture to the FMO one as follows,

$$H^{KS}_{i,j} = \sum_{\mu,\nu} c^*_{\mu,i} H^{KS}_{\mu,\nu} c_{\nu,j} .$$

Therefore, by means of the method presented in Refs. [6, 12], in the case of small overlaps $S_{i,j}$ between consecutive FMOs, the HOMO-HOMO charge-transfer-integral $t$ between the nearest neighbor GC pairs can be perturbatively obtained as

$$t_{i,j} = H_{i,j}^{KS} - \frac{1}{2}\left(S_{i,j}H_{j,j}^{KS} + H_{i,i}^{KS}S_{i,j}\right), \tag{2}$$

Here $i$ and $j$ are HOMO indexes for the molecular wire made of consecutive GC-FMOs.

The system we considered consists of a poly(GC) wire containing $N$ base pairs, contacted to left and right electrodes (figure 2). The resulting Hamiltonian can then be read as follows,

$$H = H_{DNA} + H_{leads} + H_{lead\text{-}DNA}, \tag{3}$$

with

$$H_{DNA} = \sum_{i=1}^{N}\varepsilon_i d_i^\dagger d_i + \frac{1}{2}\sum_{i\neq j}^{N}t_{i,j}\left(d_i^\dagger d_j + d_j^\dagger d_i\right),$$

$$H_{leads} = \sum_{\alpha,k}\varepsilon_{\alpha,k}c_{\alpha,k}^\dagger c_{\alpha,k},$$

$$H_{lead\text{-}DNA} = \sum_{\alpha=L,k}V_{1,k}\left(d_1^\dagger c_{\alpha,k} + h.c.\right) + \sum_{\alpha=R,k}V_{N,k}\left(d_N^\dagger c_{\alpha,k} + h.c.\right),$$

where $H_{DNA}$ is a one-dimension-chain model for DNA. $d_i^\dagger$ and $d_i$ are the electron operators on the FMO $i$. The parameters $t_{i,j}$ can be obtained from Eqs. (2), while $\varepsilon_i$ is the HOMO level of base-pair; $\alpha$ is lead index and $c$ are lead electron operators; $\varepsilon_{\alpha,k}$ are $k$ space energies in lead $\alpha$; $H_{lead\text{-}DNA}$ describes the electron hopping between DNA and leads.

The retarded/advanced Green function $G^{r/a}$ can be calculated by means of the equation-of-motion technique,

$$G^{r/a}(\omega) = \left(\omega I - H_{DNA} - \Sigma^{r/a}(\omega)\right)^{-1}, \tag{4}$$

where I is the identity operator, $\Sigma^{r/a}(\omega) = \Sigma_L^{r/a}(\omega) + \Sigma_R^{r/a}(\omega)$, while $\Sigma_{L;i,j}^{r/a}(\omega) = -i\Gamma_L\delta_{1,i}\delta_{1,j}/2$, and $\Sigma_{R;i,j}^{r/a}(\omega) = -i\Gamma_L\delta_{N,i}\delta_{N,j}/2$. $\Gamma_\alpha$ is the level-width function from lead $\alpha$.

The current can be obtained by Landauer formula,

$$I = \frac{2e}{h} \int d\omega T(\omega)[f_L(\omega) - f_R(\omega)] \tag{5}$$

$$T(\omega) = \text{Tr}\{\tilde{\Gamma}_L G^r \tilde{\Gamma}_R G^a\} \tag{6}$$

with $\tilde{\Gamma}_\alpha = i(\Sigma_\alpha^r - \Sigma_\alpha^a)$, $f_\alpha(\omega)$ is the Fermi function of lead $\alpha$, and $T(\omega)$ is the transmission function.

The poly(GC) wire stretching-twisting process can be described by in linear response as proposed by the Bustamante group[3]

$$\Delta\phi = -k\Delta d, \tag{7}$$

where $k^{-1} = g/S$, $S = 1100 \pm 200\text{pN}$ is the stretch modulus[13-16] and $g = 200 \pm 100\text{pN} \cdot \text{nm}$ is the stretch-twist coupling.[15,17-21] The stretching distance is $\Delta d = d - d_{eq}$, and the twisting angle is $\Delta\phi = \phi - \phi_{eq}$, while $d$ is the distance between the two nearest-neighbor GC pairs, $\phi$ is the angle between them, and subscript 'eq' indicates *equilibrium position* ($d_{eq} = 3.4\text{ Å}$, $\phi_{eq} = 36°$).

For every molecular mechanic configuration in the stretching-twisting process, the HOMO-HOMO charge-transfer-integral $t$ between two GC pairs is then calculated. The results for the transfer integral $t$ are shown in figure 1: Starting from the equilibrium position the distance is increased and the twist angle $\phi$ reduced following Eq. (8); $t$ is first suppressed and it then exhibits a local maximum. The physics of this process can be understood in terms of the competition between stretching and twisting. As known in the literature,[6,7] for a pure twisting process (distance $d \equiv d_{eq}$ is fixed), decreasing the angle $\phi$, implies a reduction of $|t|$ followed by a rapid increase. For a pure stretching process, $|t|$ is always exponentially suppressed as it happens in tunneling through vacuum. Therefore the full stretch-twist process can be understood as the dominance of an angle enhanced transfer integral $t$ over pure stretching after the critical value $\phi = \phi_0$ ($d = d_0 \approx 4.45\text{ Å}$). Further in the stretching process, the exponentially suppressed distance-related tunneling dominates over the increase of the $\pi$ orbital alignment in the eclipsed configuration.

The physics for this process can also be visualized by means of the density distribution of electrons in the bonding and anti-bonding orbitals of the two HOMOs (figure 3). At the equilibrium position, the overlap of $\pi$-$\pi$ bond from two GC pairs is strong enough (figure 3(a,b)). For $\phi = \phi_0$ ($d = d_0$), the HOMO of each GC pair is completely localized (figure 3(c,d)). This results in no overlap between the two HOMOs. After that, the HOMO-HOMO overlap is again increased (figure 3(e,f)), before being again suppressed due to over-distance tunneling.

Ignoring boundary effect on the level structure of DNA, we can consider site independent quantities $\varepsilon_i \equiv \varepsilon$ and $t_{i,j} \equiv t$ in rest of this work. With these parameters, we investigate the response of the current through a poly(GC) wire in dependence of the different lead-DNA couplings $\Gamma$ in the stretching-twisting process. We consider here $N = 30$ consecutive base pairs and fix the bias voltage $V_{bias}$ at -2.0 V; $\varepsilon - E_F = -1.5eV$, $E_F$ is the Fermi level at zero bias.

In DNA nanoelectronic experiments the backbone is thiol-anchored to gold electrodes, which we model with injection rate values of $\Gamma$ in the meV regime. The calculated current-distance response is shown in figure 5. Remarkably enough while reducing the electrode/molecule coupling $\Gamma$ the current signal oscillates between equal-height "on"-values and suppressed-current "off" values. The presence of such "on"-plateau in the stretching-twisting process can be understood by an analytic treatment of the two base pair limit. In this latter case, we can elaborate from Eq. (7) a formula for the transmission function in dependence of lead-DNA coupling $\Gamma$:

$$T(\omega) = \frac{\Gamma^2}{4}\left\{\frac{1}{(\omega-\varepsilon_+)^2+(\Gamma/2)^2}+\frac{1}{(\omega-\varepsilon_-)^2+(\Gamma/2)^2}-\frac{2\left[(\omega-\varepsilon_+)(\omega-\varepsilon_-)+(\Gamma/2)^2\right]}{\left[(\omega-\varepsilon_+)(\omega-\varepsilon_-)+(\Gamma/2)^2\right]^2+4t^2(\Gamma/2)^2}\right\}, \quad (8)$$

here $\varepsilon_1 = \varepsilon_2 = \varepsilon$, $t_{1,2} = t$, $\Gamma_L = \Gamma_R = \Gamma$ and $\Sigma_L = \Sigma_R = -i\Gamma/2$, $\varepsilon_\pm = \varepsilon \pm t$. For the case of $\Gamma \ll t$, at the poles $\omega = \varepsilon_\pm$, we can attain the quantum limit $T(\omega) = 1$ independently of $t$. This explains the emerging of the current "on"-plateau.

In a conclusion, we have investigated the electrical response of poly(GC) wires under a stretching-twisting mechanical process. In the overstretching regime we find local maxima for the charge-transfer-integral $t$ between two nearest-neighbor GC pairs, arising from the competition between stretching and twisting. This leads to an intermittent current response which strongly depends on the DNA-electrode coupling $\Gamma$. For those mechanical configurations, where the transfer integral $t$ is larger than $\Gamma$, a current plateau is observed. In the case of experimentally relevant small $\Gamma$'s, several equal height "on/off" switching responses are observed.

These results are supported by the independent experimental observation in the Scheer[5] and the Porath groups.[4] Our results suggest, that these finding can be understood as a competition of stretching and twisting effects on the charge transfer parameters. The poly-GC structures presented here server as a model for the more complex sequences used in experiments. However, the behavior of the $t$ parameters upon stretch and twist is very similar for other base pair combinations like A-G, AA [6-8]. Therefore, we expect for other sequences the same qualitative pictures. Of course, dynamic aspects and the effects of solvent are neglected in this model study. During dynamics, the base pairs will fluctuate around equilibrium positions, therefore proper sampling would have to calculate averages of $t$ and $\varepsilon$ along the trajectories. However, it can be expected that the behavior of the averages is qualitatively the same as for the parameters along the ideal stretching pathway as considered here. The solvent degrees of freedom lead to fluctuations in the onsite parameters. Again, averages have to be taken, which we assume to be not relevant to understand the qualitative picture.

ACKNOWLEDGMENT. We acknowledge fruitful discussions with Rafael Gutierrez. This work was funded by Deutsche Forschungsgemeinschaft (DFG) within the project CU 44/5-1 of the Priority Program SPP 1243, by the DFG project CU 44/3-2, and by the EU project IST-029192-2 "DNA-based nanoelectronic devices". Support from the Volkswagen Foundation under grant No. I/78 340 is also gratefully acknowledged.

**FIGURE CAPTIONS**

**Figure 1.** (Color online) The charge-transfer-integral *t* for the stretching-twisting process as a function of the distance between two GC pairs, *d*, and the angle $\phi$. The green dashed line represents the equilibrium position $d_{eq} = 3.4\,\text{Å}$, $\phi_{eq} = 36°$. The blue dashed-dotted line represents the suppression point ($d_0 \approx 4.45\,\text{Å}$, $\phi_0 \approx 24.6°$). The three horizontal shot-dashed gray lines are the references used for the intensity of the DNA coupling to the electrodes displayed in figure 5.

**Figure 2.** (Color online) Partition of the DNA molecular junctions for the fragment molecular orbital calculation. See text for details.

**Figure 3.** (Color online) Electron density distribution. (a) HOMO at $d_{eq} = 3.4\,\text{Å}$, $\phi_{eq} = 36°$. (b) HOMO-1 at $d_{eq} = 3.4\,\text{Å}$, $\phi_{eq} = 36°$. (c) HOMO at $d_0 \approx 4.45\,\text{Å}$, $\phi_0 \approx 24.6°$. (d) HOMO-1 at $d_0 \approx 4.45\,\text{Å}$, $\phi_0 \approx 24.6°$. (e) HOMO at $d = 5.0\,\text{Å}$, $\phi = 18.6°$. (f) HOMO-1 at $d = 5.0\,\text{Å}$, $\phi = 18.6°$.

**Figure 4.** (Color online) Coupling vs distance dependence of the transfer integral *t* at $V_{bias} = -2.0\,\text{V}$ for a 30 base pair long poly(GC). Here $\varepsilon_{GC} = -1.5\,e\text{V}$, and *d* is for the distance between two GC pairs. The green dash line is for equilibrium position ($d_{eq} = 3.4\,\text{Å}$).

**Figure 5.** (Color online) Current-distance relations along the three red (dash-dot) lines in figure 4 with same parameters. (a) $\Gamma = 9.5\,m e\text{V}$, (b) $\Gamma = 3\,m e\text{V}$, (c) $\Gamma = 1\,m e\text{V}$. The blue dash-dot lines are for the position ($d = d_0$, $\phi = \phi_0$).

**REFERENCES**

[1]   *Lecture Notes in Physics*, Cuniberti G., Fagas G., Richter K., Ed., Springer Berlin, **2005**, Vol. 68.
[2]   Porath D., Cuniberti G., and Di Felice R., *Topics in Current Chemistry* **2004**, *237*, 183.
[3]   Gore J., Bryant Z., Nollmann M., Le M. U., Cozzarelli N. R., and Bustamante C., *Nature* **2006,** *442*, 836.
[4]   Cohen H., Nogues C., Naaman R., and Porath D., *Proc. Natl. Acad. Sci. USA* **2005**, *102*, 11589.
[5]   Kang N., Erbe A., Scheer E., *New J. Phys.* **2008**, *10*, 023030.
[6]   Senthilkumar K., Grozema F. C., Guerra C. F., Bickelhaupt F. M., Lewis F. D., Berlin Y. A., Ratner M. A., and Siebbeles L. D. A., *J. Am. Chem. Soc.* **2005**, *127*, 14894.
[7]   Grozema F. C., Siebbeles L. D. A., Berlin Y. A., and Ratner M. A., *CHEMPHYSCHEM* **2002**, *3*, 536.
[8]   Felice R. D., Calzolari A., Molinari E., and Garbesi A., *Phys. Rev. B* **2001**, *65*, 045104.
[9]   Maragakis P., Barnett R. L., Kaxiras E., Elstner M., and Frauenheim T., *Phys. Rev. B* 2002, 66, 241104.
[10]  Elstner M., Porezag D., Jungnickel G., Elsner J., Haugk M., Frauenheim T., Suhai S., and Seifert G., *Phys. Rev. B* **1998**, *58*, 7260.
[11]  Frauenheim T., Seifert G., Elstner M., Hajnal Z., Jungnickel G., Porezag D., Suhai S., and Scholz R., *Phys. Stat. Sol. (b)* **2000**, *217*, 41.
[12]  Newton M. D., *Chem. Rev.* **1991**, *91*, 767.
[13]  Bustamante C., Smith S. B., Liphardt J., and Smith D., *Curr. Opin. Struct. Biol.* **2000**, *10*, 279.
[14]  Bryant Z., Stone M. D., Gore J., Smith S. B., Cozzarelli N. R., and Bustamante C., *Nature* **2003**, *424*, 338.
[15]  Smith S. B., Cui Y., and Bustamante C., *Science* **1996**, *271*, 795.
[16]  Wang M. D., Yin H., Landick R., Gelles J., and Block S. M., *Biophys. J.* **1997**, *72*, 1335.
[17]  Strick T. R., Allemand J.-F., Bensimon D., Bensimon A., and Croquette V., *Science* **1996**, *271*, 1835.
[18]  Marko J. F., *Europhys. Lett.* **1997**, *38*, 183.
[19]  Kamien R. D., Lubensky T. C., Nelson P., and O'Hern C. S., *Europhys. Lett.* **1997**, *38*, 237.
[20]  Moroz J. D. and Nelson P., *Macromolecules* **1998**, *31*, 6333.
[21]  Cluzel P., Lebrun A., Heller C., Lavery R., Viovy J.-L., Chatenay D., and Caron F., *Science* **1996**, *271*, 792.

# Figures

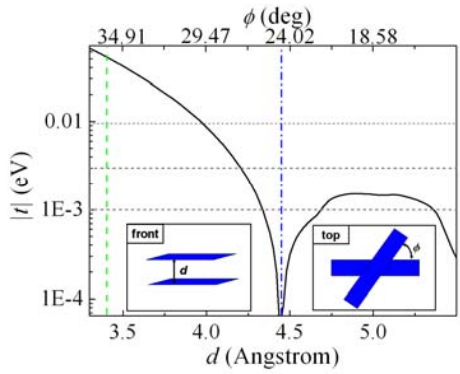

Figure 1

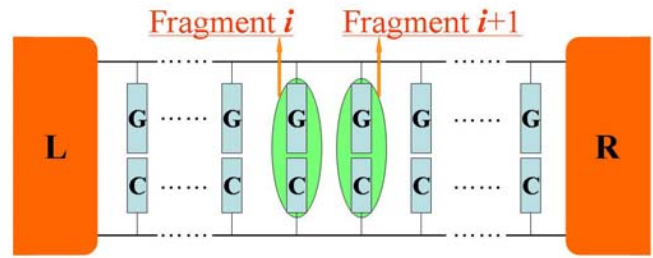

Figure 2

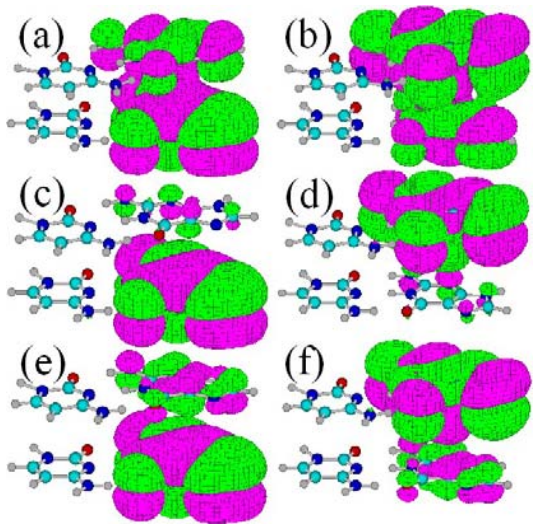

Figure 3

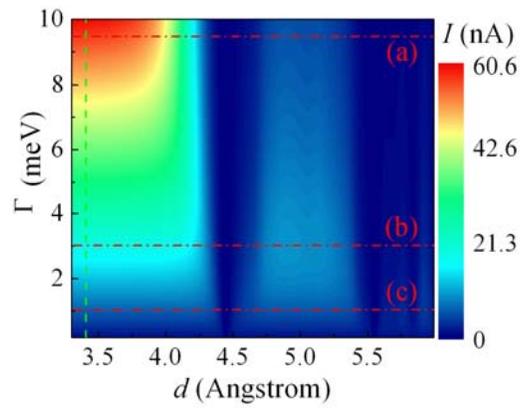

Figure 4

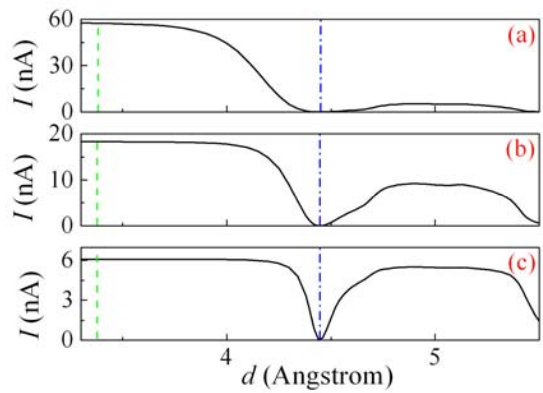

Figure 5